# Quantum-enhanced laser phase noise filter

*Ruixin Li[1], Nanjing Jiao[1], Bingnan An[1], Yajun Wang[1,2]\*, Shaoping Shi[1,2], Long Tian[1,2], Wei Li[1,2], Yaohui Zheng[1,2]\*\**

[1]*State Key Laboratory of Quantum Optics Technologies and Devices, Institute of Opto-Electronics, Shanxi University, Taiyuan 030006, China*
[2]*Collaborative Innovation Center of Extreme Optics, Shanxi University, Taiyuan, Shanxi 030006, China*

\**e-mail: YJWangsxu@sxu.edu.cn*
\*\**e-mail: yhzheng@sxu.edu.cn*

## Abstract

Quantum noise is the fundamental limit of laser phase noise filter. We cannot realize the effective quantum-enhanced phase noise suppression through simply utilizing amplitude noise suppression scheme. Here, we present the first experimental demonstration of a quantum-enhanced laser phase noise filter, achieved by employing a noise ellipse rotation phase noise readout technique combined with an excess amplitude noise suppression scheme. We address the primary limitations in the extracting of laser phase noise, and make the quantum enhancement via squeezed vacuum injection feasible. A maximum of 5 dB quantum-enhanced phase noise suppression is realized across the Fourier frequencies from 5 kHz to 60 kHz. The demonstration unlocks the application of squeezed vacuum state in laser phase noise suppression.

*Introduction.–* Laser noise, as a frequently encountered phenomenon [1], limits the resolution of many optical measurement [2-4] and sensing apparatuses [5-9], as well as the minimum detectable signal [10-14]. Therefore, laser noise suppression becomes a basic and essential building block for unprecedented sensitivity, such as gravitational waves observatory that needs the ability to measure tiny signal equivalent to a trillionth of a wavelength [15-17]. In addition, for trapping of cold atoms, laser with lower noise is critical to ensure lower atom heating rates which, in turn, maximizes the experiment time [18,19]. The laser noise, including amplitude and phase noises, originates from quantum noise due to spontaneous emission of the gain medium and optical losses, as well as technical noise that can be completely suppressed [20-22]. Infinite striving to be the best sensitivity drives the relentless innovation and breakthrough of laser noise suppression [23,24].

Feedback control is a popular technical route for suppressing the laser noise,

particularly at low-frequency band [25-29]. Laser amplitude noise can be actively controlled by directly detecting a fraction of the laser field with a photodetector (PD) and then compared with a reference standard to generate an error signal. One of the key factors that restricts the achievable noise suppression of the feedback loop is signal to noise ratio (SNR) of the detection process. Although tremendous efforts that cover an innovative PD with low-noise and high saturation power [21,22], and an elegant power detection scheme called optical AC coupling [30,31], have been made, the in-loop noise of the feedback loop cannot break the shot noise limit (SNL). The laser amplitude stabilization scheme beyond the SNL by the employment of squeezed vacuum states of light was proposed as early as 1987 [32]. Owing to a reduced noise in one quadrature of squeezed vacuum field, the SNR of the measured amplitude fluctuation can be reduced without increasing the detected power. More than 30 years later, laser amplitude stabilization beyond the SNL was experimentally demonstrated [33,34].

Unfortunately, the laser phase noise cannot be directly read out from the laser output [35-38], making the squeezing-enhanced phase noise suppression face great challenge. Commonly, two types of phase quadrature readout technologies are utilized to extract the phase noise. One is a heterodyne detection technique [37,38], accompanying with a large number of technical (optical, electronic and mechanical) noises [28,29], especially the 3 dB quantum noise penalty will restrict the application of quantum-enhanced noise suppression [39-41]. The other one is the direct optical readout technique [35,42-44], converting the phase quadrature to an amplitude one, which enables to be directly readout by a PD. It usually suffers from a serious technical (optical/amplitude) noise at the low frequency, especially in a low efficiency readout technique [45]. But it has the advantage of single technical noise source and insignificant quantum noise penalty, similar to the amplitude noise suppression technique, which makes the quantum-enhanced noise suppression regime become possible. Nevertheless, we must have a deep understanding of the physical mechanism in noise readout and suppression processes, and set the stage for implementing the quantum-enhanced regime.

In this Letter, we report for the first time on the experimental realization of a squeezing-enhanced laser phase noise suppression. Exploiting a rotation of noise ellipse scheme, we partly extracted the phase noise reflected from a one-ended cavity. By the employment of a pre-stabilized scheme, the amplitude noise was suppressed to near the SNL among the kHz to MHz frequency band, significantly mitigating the influence of the amplitude noise on error signal extraction. At last, we measure a quantum-enhanced laser phase noise in the Fourier frequency range of 5 to 60 kHz with the assistance of a squeezed vacuum state, corresponding to a maximum noise reduction enhancement of 5 dB. The experimental results agree well with the theoretical one. The demonstration unlocks the potential of further lowering laser phase noise through quantum-enhanced technique.

*Principle of quantum-enhanced phase noise filter.*–The fundamental principle of our quantum-enhanced phase noise filter is illustrated in Fig. 1. To make the physical mechanism clearer, we initiate the quantum-enhanced scheme from amplitude noise suppression, in which the noise can be directly detected by a low noise PD, as shown in Fig. 1(a). As the gain of servo loop goes to infinity, the vacuum noise, from the input port of beam splitter (BS) coupled to the in-loop sensing beam, becomes the fundamental limit of amplitude noise suppression [32,33]. With audio-squeezed vacuum injection, the sub-SNL in-loop noise can be reached [33]. Inspiring by the ideology of Fig. 1(a), we develop the first quantum-enhanced phase noise filter based on a direct optical readout unit, by employing a rotation of noise ellipse technique, as shown in Fig. 1(b). In virtue of a phase-to-amplitude converter, the phase quadrature of the laser beam partly converts into an amplitude one, vice versa, and the phase noise can be directly measured with a PD, carrying out the active feedback control. However, the detected fluctuations contain the joint contributions of the phase and amplitude components, due to a small rotation angle of the noise ellipse. The excess amplitude noise contributes to the residual technical noise, and reduces the efficiency of phase noise suppression as well as the quantum-enhanced performance, which should be in advance suppressed as low as possible before squeezing injection. Therefore, we apply a shot-noise-limited amplitude noise filter to make the phase noise extraction more accurate, as illustrated in Fig. 1 (b) and (c).

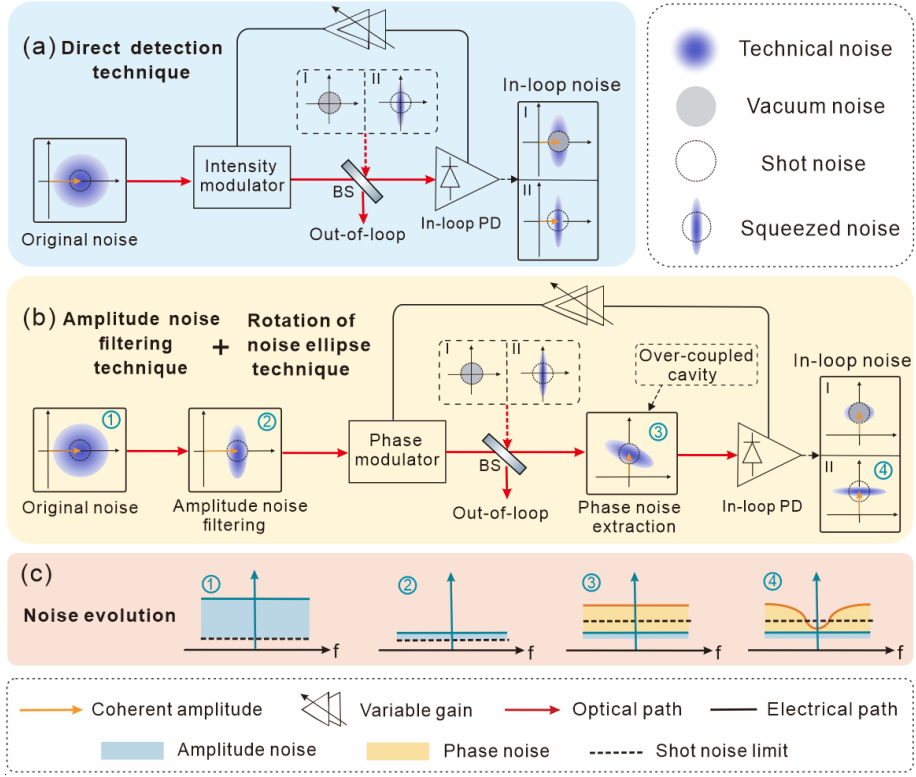

FIG. 1 Physical mechanism of quantum-enhanced laser noise filter with active feedback control loop. (a) amplitude noise suppression by an active feedback control with directly detecting the laser field; (b) rotation of noise ellipse technique for quantum-enhanced phase noise suppression in combination with amplitude noise pre-stabilization; (c) noise evolution of the amplitude quadrature in the coherent amplitude space (expressed in relative noise form), corresponding to the configuration shown in (b). BS: Beam splitter; PD: Photodiode.

We begin the quantum-enhanced laser phase noise suppression analysis with shot-noise-limited amplitude noise performance (position ②). The laser field firstly transmits through a phase modulator, after which a portion of the field passes through the BS with a transmission of $t$ and reflectivity of $r$. A part of the phase quadrature transmitted from the BS is readout by a phase noise ellipse rotation technique (an over-coupled cavity, OCC, position ③) [35,42-43] and a PD, converting the phase quadrature into amplitude one and acting as the feedback signal of the in-loop. The amplitude quadrature of the in-loop laser field after OCC can be expressed as [35,42-43]

$$X_{in-loop}^d = X_{in-loop}\cos\theta_1 + Y_{in-loop}\sin\theta_1 \quad (1)$$

where $\theta_1$ represents the relative phase between the coherent and squeezed vacuum fields, determined by the cavity detuning, and relating to the noise ellipse rotation angle $\theta_1$. It implies that the coherent field is rotated with 90° by the OCC, but the quadrature changes an angle of $\theta_1$ with respect to the coherent field (position ③). An amplitude squeezed vacuum is coupled with the input field at the BS, which also experiences the same rotation angle after OCC. To realize an optimum quantum-enhanced noise suppression, the amplitude $X_{in-loop}$ and phase $Y_{in-loop}$ quadratures of the coupled laser field before OCC can be written as

$$X_{in-loop} = \sqrt{t}X_{in} + \sqrt{r}X_S\cos\theta_1 + \sqrt{r}Y_S\sin\theta_1$$
$$Y_{in-loop} = \sqrt{t}Y_{in} - \sqrt{r}Y_S\cos\theta_1 + \sqrt{r}X_S\sin\theta_1 \quad (2)$$

where $X_{in}$, $Y_{in}$ and $X_S$, $Y_S$ are the amplitude and phase quadratures of the input and squeezed vacuum fields, respectively. The minus in the second term of $Y_{in-loop}$ is applied to cancel the anti-squeezed quadrature, which can be achieved by locking the relative phase between the squeezed vacuum and input laser field at the BS to $-\theta_1$. Substituting Eq. (2) into Eq. (1), we obtain

$$X_{in-loop}^d = \sqrt{t}X_{in}\cos\theta_1 + \sqrt{t}Y_{in}\sin\theta_1 + \sqrt{r}X_S \quad (3)$$

Subsequently, we replace the noise variance as $S = X^2$, and the phase quadrature rotates with $\theta_1 \cong \omega/\kappa_1$. Attributing to the phase-to-amplitude noise conversion process, the PD can directly measure a part of the phase noise of the input field. Hence, we interpret the phase noise into the relative intensity noise (RIN) representation [46]

$$RIN_{in-loop}^2 = \frac{2h\nu}{P_{in-loop}}(tS_{in}^X + rS_S^X + t\frac{\omega^2}{\kappa_1^2}S_{in}^Y) \quad (4)$$

where, $\nu$ is the laser frequency, $\omega = 2\pi f$, $f$ is the analysis frequency, $\kappa_1$ is the linewidth of the OCC, and $h$ is the Planck's constant. Decomposing $S_{in}$ into technical noise $S_{tech}$ and quantum (shot) noise $S_{sn}$, we rewrite $S_{in}^{X,Y} = S_{tech}^{X,Y} + S_{sn}^{X,Y}$, $2h\nu S_{X,Y}/P_{in} = RIN_{X,Y}^2$, and $2h\nu/P_{in-loop} = RSN_{in-loop}^2$. Under the experimental conditions of $t \ll 1$ and $S_{tech}^Y \gg S_{sn}^Y$, Eq. (4) is simplified as

$$RIN_{in-loop}^2 \cong RIN_X^2 + \frac{\omega^2}{\kappa_1^2}RIN_Y^2 + RSN_{in-loop}^2 \cdot S_S^X \quad (5)$$

The first term of Eq. (5) corresponds to the amplitude noise of the laser field at position ②. The second term represents the converted phase noise at position ③. The third term is determined by the shot noise measured at PD and the squeezed vacuum (position ④).

Apparently, the first two terms contribute the technical noise of the feedback control loop, which should be suppressed far below the noise floor of the third term to guarantee the action of quantum enhancement.

With the same noise readout method, the measured RIN of the phase noise in the out-of-loop can be expressed as

$$RIN_{out-of-loop}^2 \cong RIN_X^2 + \frac{\omega^2}{\kappa_2^2} RIN_Y^2 \tag{6}$$

where, $\kappa_2$ denotes the linewidth of the impedance-matched cavity (IMC). Considering the assistance of feedback control, the out-of-loop *RIN* evolves to

$$RIN_{out-of-loop}^2 = RIN_X^2 + \frac{\omega^2}{\kappa_2^2} \left( \frac{RIN_Y^2}{|1-\sqrt{t}G(\omega)|^2} + \frac{RSN_{in-loop}^2 \cdot S_S^X}{r} \right) \tag{7}$$

where $G(\omega)$ is the feedback gain at frequency $\omega$. The first term of Eq. (7) represents the amplitude noise of the optical field, which sets the lower bound of the SNR for phase noise measurement. The second term is the technical noise relating to the residual excess phase noise after feedbacking. The last term is imprinted by the combing of the in-loop shot noise and squeezed vacuum enhancement.

*Experimental setup.–*Our experimental setup of quantum-enhanced phase noise suppression is shown in Fig. 2. A single-frequency fiber laser at 1550 nm with the output power of 1 W is employed as the laser source that serves for downstream experiment. Its output passes through an acousto-optic modulator (AOM) with a frequency-shift of 110 MHz, diffraction efficiency of 50%, which allows for a fast phase modulation of the output field within 200 kHz bandwidth. The major fraction of the light field about 500 mW traverses through the optical isolator (ISO), and pours into the second harmonic generator (SHG) to convert the incoming field to a wavelength of 775 nm. The output field provides the pump field for the parametric down-conversion process in a doubly resonant optical parametric oscillator (OPO). The detailed description of the design and operation of the SHG and OPO systems, including their control schemes, can be found in Refs. [47–49]. By adjusting the pump power using a half-wave plate (HWP2) and a polarizing beam splitter (PBS2), we set the pump field to 10 mW before injecting it into the OPO, ensuring that the OPO operates well below its oscillation threshold of 20 mW. The squeezed vacuum states of light at 1550 nm, generated by the OPO, are separated from the pump light by a dichroic beam splitter (DBS).

It is worth noting that the SHG also serves as an amplitude noise suppressor, whose mechanism relies on the frequency conversion process inducing a nonlinear relation between the intracavity circulating and the reflected powers of the fundamental wave [47,50]. The maximum amplitude noise reduction is achieved when the slope of this nonlinear transfer function approaches zero, which is suppressed to near the SNL at frequency band of kHz-MHz under the conversion efficiency of 70% [47,50]. It efficiently eliminates the influence of the amplitude noise on the phase noise suppression in the downstream experiment (position ①→②). The reflected fundamental wave with a power of 100 mW was firstly attenuated to 5 mW by a HWP1 and a PBS1. Then it is poured into a BS with a power reflectivity of $r$=99%, which splits the laser field into in-loop and out-of-loop beams. Approximately 50 μW of laser field is injected into an OCC, which serves as the phase-to-amplitude quadrature convertor for the in-loop laser field (position ③). Its input coupler has a power reflectivity of 95%, and that of the output one is 98.5%. The third one is a concave mirror with high reflectivity of better than 99.95%. It results in a cavity linewidth of 7.5 MHz [51]. The design ensures enough power extracting from the transmitted laser for stable locking of the detuned OCC with fringe-side locking technique [44,51]. We can extract the phase noise with in-loop PD at the reflected port of the OCC, whose photocurrent is fed back by a PID controller (Vescent, model: D2-125) on the upstream AOM to implement an active control of the phase noise. However, the transmittance of the OCC induces an additional loss for the squeezed vacuum field coupling, which will reduce the quantum-enhanced noise level. To balance the noise conversion efficiency in the reflected port and loss in transmission, we lock the OCC to approximately three times the half-detuning status, with only 1.6% additional loss introduction. In this case, the quadrature rotation angle becomes approximately $\theta_1 \cong 0.1 \cdot \omega/\kappa_1$ Therefore, to ensure an optimum quantum-enhanced noise reduction, the relative phase between the squeezed vacuum and input field at the BS is accurately locked to -$\theta_1$ with a coherent control technique [52]. The out-of-loop employs an independent three-mirror half-detuned IMC to characterize the phase noise, whose principle is the same as the OCC, but has a linewidth of 6.8 MHz due to a 97% power reflectivity of the two plane mirrors.

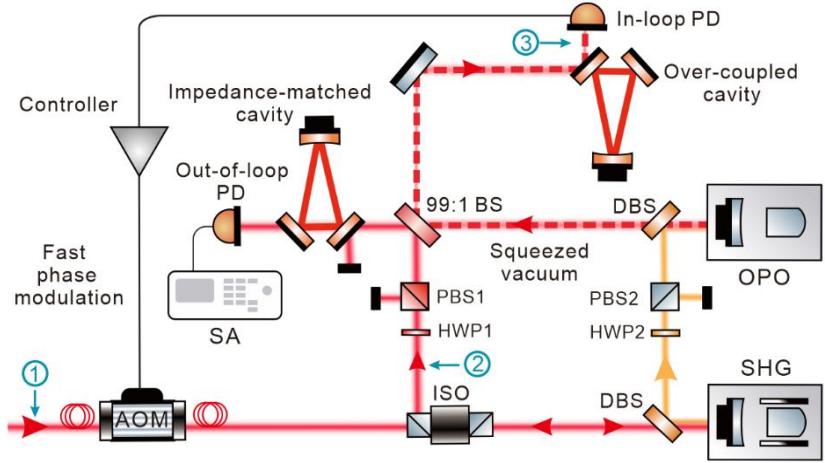

FIG. 2 Experimental schematic diagram of the quantum-enhanced phase noise filter. AOM, acousto-optic modulator; ISO, isolator; SHG, second harmonic generation; DBS, dichroic beam splitter; OPO, optical parametric oscillator; HWP: half-wave plate; PBS, polarization beam splitter; PD, photodetector; 99:1 BS, 99:1 beam splitter; SA, spectrum analyzer.

*Experimental results and discussion.*–The free-running frequency noise in our experiment is directly measured by the IMC under the optimized amplitude noise stabilization, as shown in trace (a) of Fig. 3. The fundamental wave RIN is suppressed down to -157 dB/Hz through the frequency-doubling process[47]. Consequently, the residual excess amplitude noise is assumed to always keep as -157 dB/Hz (14.1 dB below the in-loop's SNL) at the measurement frequency shown in trace (f), due to the fact that the quadrature is only rotated by the IMC with a small angle inferred by Eq. (3) or (4). The electronic noise is directly measured to be approximately -160 dB/Hz, shown in trace (g). The low electronic and residual excess amplitude noises contributes to the boundary of technical noise floor of the system for quantum-enhanced noise suppression. With feedback control loop operation, the in-loop noise floor is actively suppressed as shown in trace (e), corresponding to the second term of Eq. (7).

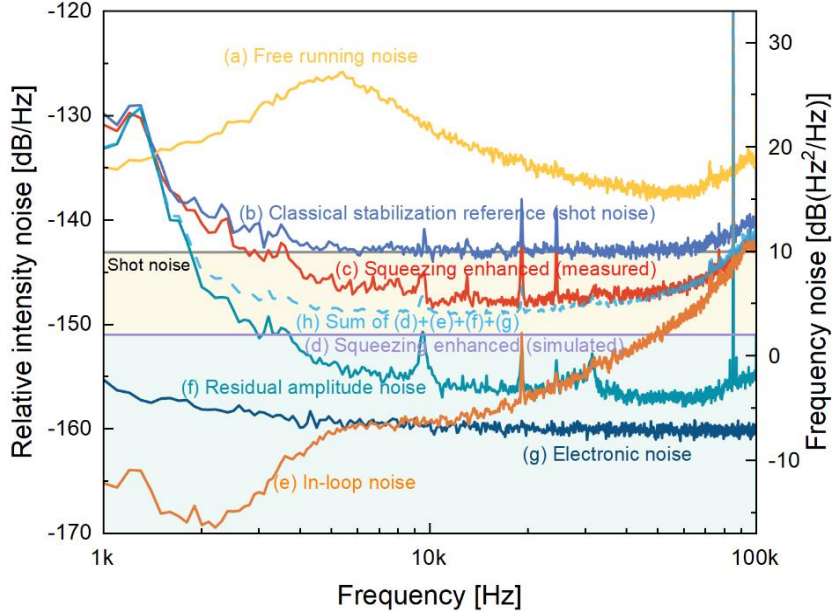

FIG. 3 Experimental results and noise budget of the quantum-enhanced laser phase filter. Trace (a): Measured free-running laser frequency noise. Trace (b): Measured out-of-loop laser frequency noise corresponding to the shot noise level of the 50 μW in-loop detected power. This is the classical stabilization reference. Trace (c): Nonclassical laser frequency noise reduction. Between 5-60 kHz an average of 5 dB below the shot noise was achieved. Trace (d): Simulated squeezing enhancement. Trace (e): In-loop noise during squeezing-enhanced electronic loop operation. Trace (f): Residual amplitude noise. This is a direct measurement results before phase-to-amplitude conversion. Trace (g): Electronic noise of the in-loop PD. Trace (h): Uncorrelated sum of limiting noise sources (d),(e),(f),(g). All measurement were performed with a spectrum analyzer (R&S FSW) at Fourier frequencies from 1 kHz to 100 kHz with RBW 100 Hz.

Without squeezed light injected, the out-of-loop laser frequency noise reduction is limited by the in-loop laser shot noise level of -142.9 dB/Hz as shown in trace (b). We employ an amplitude squeezed vacuum state that couples with the laser beam ② on a 99:1 BS with mode matching efficiency of 98.5±0.2% and a locked relative phase of $\theta_1 \approx 0$, the quantum-enhanced suppression of the phase noise is shown by trace (c), where the frequency noise spectrum exhibits a maximum noise reduction of 5 dB between 5-60 kHz, comparing with the classical noise suppression level of trace (b). In general, the frequency noise power spectral density is given by $S_\nu(f) = f^2 S_\varphi(f) = f^2 RIN_Y^2$, where $S_\varphi(f)$ represents the power spectral density of phase noise. It is evident that the frequency noise spectra of the in-loop and out-of-loop are independent of cavity linewidth, which are identical. But, the out-of-loop RIN shown as the traces (a)-(c) are normalized by a factor of $(0.1 \cdot \kappa_2)^2/\kappa_1^2$, corresponding to the different measurement linewidth of the in-loop cavity.

The discrepancy between the generated squeezed vacuum and quantum enhancement attributes to the optical loss, phase fluctuations and noise cross-coupling, which are summarized in Table 1. Individual system loss effects are also shown in the Table1, corresponding to a squeezing strength degradation of 1.9 dB [53,54]. In our system, three phase locking loops are employed to introduce a total phase fluctuation of 20±0.9 mrad, leading to a 0.6 dB squeezing noise deterioration. If only considering the optical losses and phase fluctuations of squeezed state generation, the theoretically achievable quantum enhancement is 8.1 dB, as indicated by Eq. (7) and the dashed line in trace (d). In addition, the feedback process is also accompanied by the noise-coupling of the electronic noise of PD (trace (g)), residual excess amplitude noise (trace (f)) and in-loop frequency noise (trace (e)). All these noise-coupling factors contribute to a 2.2 dB degradation of quantum enhancement, limiting the theoretical quantum enhancement to 5.9 dB (trace (h)). Finally, the non-ideal servo loop, including the finite gain and zero drift and so on, also deteriorates the measured result, ultimately to 5 dB quantum enhancement. The higher noise coupling below 10 kHz is mainly raised by the cavity detuning of the SHG within its locking bandwidth, which is induced by the modulation of upstream AOM under phase noise feedbacking process. We expect to extend the bandwidth of the phase noise suppression through technology innovations in the photodetector, actuator and servo loop [34].

TABLE. 1 Noises and system errors budget from the optical loss, phase fluctuation, and noise cross-coupling

| Source of optical loss | Value (%) | System loss (dB) |
|---|---|---|
| OPO escape efficiency | 97±0.5 | |
| Efficiency of interference | 98.5±0.2 | |
| Quantum efficiency of photodiodes | 99±0.2 | 1.9 |
| Over-coupled cavity | 98.4±0.5 | |
| Laser propagation efficiency | 96±0.5 | |
| Total efficiency | 88±0.8 | |
| **Source of phase fluctuation** | **Value (mrad)** | **System loss (dB)** |
| OPO cavity length | 2±0.3 | |
| Relative phase between squeezed and frequency-shifted light | 7±0.5 | 0.6 |
| Relative phase of squeezed and local oscillator | 11±0.6 | |
| Total phase fluctuation | 20±0.9 | |
| **Source of noise-coupling** | **Value (%)@8 kHz** | **System loss (dB)** |
| Electronic noise | 2.3±0.1 | |
| Residual laser excess amplitude noise | 6.2±0.2 | 2.2 |
| In-loop frequency noise | 2.1±0.5 | |
| Total cross-coupling | 10.6±0.8 | |

*Conclusion and Outlook.*–We have demonstrated the first experimental realization of quantum-enhanced laser phase noise filter. By employing a hybrid noise reduction scheme, the technical noises of the amplitude and phase quadratures of the laser field are suppressed below the SNL of the detected laser field of the in-loop. The design with an optical direct phase noise readout method provides the fundamental conditions for squeezing enhancement. By the assisting of a 10.6 dB squeezed vacuum state, we reported a maximum of 5 dB quantum-enhanced phase noise suppression at Fourier frequencies between 5 kHz and 60 kHz. The experimental results agree well with the theoretical prediction. The methodology unlocks the application of squeezed vacuum state in laser phase noise reduction, and offers a promising strategy for advancing quantum metrology and precision measurement.

We acknowledge financial support from the National Natural Science Foundation of China (NSFC) (Grants No. 62225504, No. U22A6003, No. 62027821, No. 62375162), National Key Research and Development Program of China (No. 2024YFF0726401).